\documentclass[pra,onecolumn,11
pt,aps,nofootinbib,superscriptaddress,showkeys,showpacs,longbibliography]{revtex4-1}
\usepackage{amsmath,amsfonts,amssymb,amstext,amscd,amsthm,dsfont}
\usepackage{physics}
\usepackage{color}
\usepackage{soul,xcolor}
\usepackage{gensymb}
\usepackage[normalem]{ulem}

\usepackage{graphicx}
\usepackage{bbold}
\usepackage{braket}
\usepackage{placeins}
\renewcommand{\max}{{\rm max}}

\begin{document}
	\title{Non-Markovian evolution: a quantum walk perspective}

	\author{N. Pradeep Kumar}
	\email{pradeepnpk@imsc.res.in}
	\affiliation{The Institute of Mathematical Sciences, C. I. T. Campus, Taramani, Chennai 600113, India}
	
	\author{Subhashish Banerjee}
	\email{subhashish@iitj.ac.in}
	\affiliation{Indian Institute of Technology Jodhpur, Jodhpur 342011, India}
	
	\author{R. Srikanth}
	\email{srik@poornaprajna.org}
	\affiliation{Poornaprajna Institute of Scientific Research,
		Bangalore- 560 080, India}
	
	\author{Vinayak Jagadish} 
	\email{JagadishV@ukzn.ac.za}
	
	\author{Francesco Petruccione}
	\email{petruccione@ukzn.ac.za}
	
	\affiliation{Quantum Research Group, School  of Chemistry and Physics,
		University of KwaZulu-Natal, Durban 4001, South Africa}\affiliation{ National
		Institute  for Theoretical  Physics  (NITheP), KwaZulu-Natal,  South
		Africa}
	
	\begin{abstract}
		Quantum  non-Markovianity  of  a   quantum  noisy  channel  is
	typically identified  with information backflow,
	or,  more generally,  with departure  of the  intermediate map
	from complete positivity. But  here, we also indicate 
	certain non-Markovian channels that  can't be witnessed by the
	CP-divisibility     criterion.    In     complex    systems,
	non-Markovianity becomes more involved on account of subsystem
	dynamics.   Here  we  study various  facets  of  non-Markovian
	evolution,  in  the  context  of coined  quantum  walks,  with
	particular   stress  on   disambiguating   the  internal   vs.
	environmental  contributions to  non-Markovian backflow.   For
	the  above problem  of  disambiguation, we  present a  general
	power-spectral technique based on a distinguishability measure
	such as  trace-distance or correlation measure  such as mutual
	information.   We   also  study  various  facets   of  quantum
	correlations  in  the  transition from  quantum  to  classical
	random  walks,   under  the  considered   non-Markovian  noise
	models. The potential for the  application of this analysis to
	the  quantum  statistical  dynamics   of  complex  systems  is
	indicated.
	\end{abstract}
	
	\pacs{03.65.Ud, 03.67.-a}
	\maketitle 
	\section{Introduction}
\setcounter{equation}{0}
The  interaction  between quantum  system  and  various subsystems  is
inevitable  for  achieving   complex  quantum  information  processing
tasks. The subsystems could  either be some uncontrollable environment
that           leads            to           dissipation           and
decoherence \cite{breuer2002theory,chandrashekar2007symmetries,banerjee2008symmetry,Phasediff,decwithoutdiss}
or  it could  be used  to implement  sophisticated techniques  such as
reservoir  engineering, which  enhances  the  stability of  peripheral
tasks  such  as  quantum  state preparation  and  readout  in  quantum
computers    \cite{murch2012reservoir,err1,shankar2013reservoir,err2}.
  While the  study of  open system  dynamics presumes  an
environment that  is largely inaccessible, a  greater understanding of
the  influence   of  the  environment   on  the  system   dynamics  is
indispensable for the advancement of quantum technologies. \\

A huge body  of works on open system dynamics  has concentrated on the
Markovain approximation, where the  environmental time-scales are much
shorter  than  the  system  time-scales  \cite{RHP14,VA17}.   However,
current  advancement   in  experimental  techniques  allows   for  the
possibility  of  going  beyond  the  standard  Markovian  regime  into
non-Markovian regime where the reservoir effectuates memory effects in
the system  dynamics. 

 In particular, an example of a non-Markovian experimental
realization  would  be \cite{RHHH01,THC+07,HCH+10},  a  nanomechanical
oscillator interacting  with a  BEC in  a double-well  potential.  The
atoms of the  condensate are confined in a double  well and can tunnel
from one side of the potential to the other, depending on the position
of the oscillator.   If the condensate is taken to  be the environment
for the  oscillator, highly non-Markovian  effects appear that  can be
observed   in    the   non-exponential   decay   of    the   oscillator
coherences \cite{BAS+11,ABS14}.  In a recent experiment
\cite{GTP+15}  it was shown that the  spectral density of the  environment in
Quantum  Brownian motion  (see below)  in an  optomechanical resonator
coupled to a heat bath, is highly non-Ohmic and produces non-Markovian
dynamics in the resonator.

Quantum Brownian  motion is a paradigm  model of
quantum statistical  mechanics and has  provided a fertile  ground for
studies   related   to   non-Markovian   phenomena,   antedating   the
corresponding developments in quantum information
\cite{grabert1988quantum,hu1993quantum,banerjee2003general,banerjee2000quantum}.

 From  a   quantum  information  theoretic  perspective,
non-Markoviantity  is   now  usually  defined  by   the  condition  of
information   backflow,   by   which   is  meant   the   increase   of
distinguishability  with  time  between   any  two  given  states,  as
witnessed by measures like trace distance
\cite{breuer2009measure,breuer2016colloquium,rajagopal2010kraus};
or by the weaker condition of  CP-divisibility, by which is meant that
the time-evolution generated  by the dynamical maps  cannot be divided
into    intermediate    maps     that    are    completely    positive
(CP-Divisibility) \cite{RHP10,RHP14}.  It is known that  for invertible
maps, these two conditions are equivalent.

In general not  all given  dynamics strictly  satisfy both  the above
conditions  for it  to be  termed  non-Markovian  \cite{chruscinski2011divisiblity}. CP divisibility is known to imply a lack of information backflow, with the converse also holding true for for invertible maps. Though  efforts  towards
connecting  the divisibility  and information  flow pictures  has been
undertaken in
\cite{chruscinski2011divisiblity,bogna2017constructive},  the  general
conditions  under which  both  these methods  coincide  for any  given
non-Markovian dynamics is still an open question.  In \cite{Dariusz2018}, it is shown that even if the map is non-invertible, lack of information backflow implies a completely positive propagator, which may, however not be trace preserving.

In this work,  we wish to demonstrate the occurrence  of backflow in a
discrete-time  quantum walk  (DTQW)  system, the  quantum analogue  of
``classical random walk'' (CRW). Quantum walk is a ubiquitous platform
for studying diverse features such as symmetries, quantum to classical
transition \cite{banerjee2008symmetry}. The simplest  instance of  DTQW is
that  of  a quantum  system  with  two  levels  translating on  a  one
dimensional             discrete            position             space
\cite{ambainis2001one,chandrashekar2010relativistic}, a topology which
we use  in this work. What  makes DTQW especially interesting  is that
even in the noiseless case, the reduced dynamics of the coin manifests
non-Markovian recurrence behavior due to interaction with the position
degree of freedom \cite{Hinarejos2014}.   This could be also envisaged
in more  complex systems which  may possess an  inherent non-Markovian
feature    because    of    interaction    among    subsystems.     In
\cite{Banerjee2017NMQW}, a  method was  proposed to  disambiguate this
internal   source    from   environmental    non-Markovianity,   whose
applicability would extend beyond the  model chosen here.  The various
noisy channels we  consider are a local  dephasing non-Markovian noise
model \cite{daffer2004depolarizing}  modeled on the  random telegraph
noise (RTN)  process \cite{rice1944}, the  modified Ornstein-Uhlenbeck
(OU) \cite{uhlenbeck1930theory,yu2010entanglement}  and the  power law
noise  (PLN)  \cite{KJ11}.  Backflow  can  also  be witnessed  by  the
revival of  various facets of  quantum correlations in  the transition
from  quantum to  classical \cite{chakrabarty2010study}  random walks,
under the considered non-Markovian noise models.

We begin with a brief description of the DTQW as well as the different
non-Markovian noise  models used in this  work. This is followed  by a
study  of the  variance of  the quantum walk (QW),  evolving under  the influence  of
different noises.  Signatures of non-Markovianity are probed next by making use of trace distance  and mutual  information.  This  is followed  by a discussion  of  the  recently   developed  power  spectrum  method \cite{Banerjee2017NMQW} to
disambiguate the different sources  of non-Markovianity. The recurring
theme of  this work, i.e., to  develop insights into the  non Markovian
behavior is  further developed by  studying various facets  of quantum
correlations between the coin and position states of the DTQW. We then
make our conclusions.

\section{Brief introduction to DTQW and non-Markovian noise models}

\subsection{DTQW: A brief introduction}
The formalism of  DTQW requires two essential  ingredients, the $coin$
and the $walker$ \cite{kempe2003quantum}, which describes the internal
and external  degrees of  freedom of  the particle,  respectively. The
coin  degree  of freedom  is  spanned  by  the basis  set  $\{\ket{0},
\ket{1}\}$   $\in   \mathcal{H}_c$   and  the   walker   or   position
translational degree of  freedom is spanned by  the basis $\{\ket{i}\}
\in \mathcal{H}_p$ with $i \in  \mathbb{Z}$ representing the number of
lattice sites  available to the walker.  The state of total  system is
described  by the  Hilbert space  $\mathcal{H}_w=\mathcal{H}_c \otimes
\mathcal{H}_p$.

To implement the  DTQW, we initialize a  quantum state $\ket{\psi(0)}$
specified by two parameters $\delta, \eta$,
\begin{equation}
\ket{\psi(\delta, \eta)} = 
(\textmd{cos}(\delta)\ket{0} + e^{-i\eta}\textmd{sin}(\delta)\ket{1})
\ket{x}
\end{equation}
where  $x  \in  \{-n,-(n-1),\cdots,0,1,\cdots,  n\}$,  and  the  first
register refers  to the  coin degree  of freedom,  which lives  in the
two-dimensional Hilbert space $\mathcal{H}_C$, and the second register
refers  to  the  position  degree  of  freedom,  which  lives  in  the
$(2n+1)$-dimensional  Hilbert space  $\mathcal{H}_P$.  Here,  we shall
set    the   initial    walker   state    $\ket{\psi(t=0)}$   to    be
$\ket{\psi(\delta=\frac{\pi}{4}, \eta=0)}\ket{x=0}$. 
 
The quantum walker is then evolved  using the coin and shift (position
translation)  operators. The  action of  coin operator  determines the
direction  in which  the  particle  moves. The  general  form of  coin
operator is that of a rotation matrix $U(2)$, in two dimensional Hilbert space, given by
\begin{equation}
\hat{C}=
\begin{bmatrix}
\cos\theta_c & \sin\theta_c \\ \label{SU2}
\sin\theta_c & -\cos\theta_c  
\end{bmatrix}.               
\end{equation}
For $\theta_c=45\degree$, $\hat{C}$ reduces  to Hadamard operator, the
quantum  walk  with  such  a  coin is  also  called  Hadamard  quantum
walk \cite{ambainis2001one}.

The  shift operator  that translates  the particle  to either  left or
right is conditioned on the outcome of the coin operator.  The general
form of the shift operator is given as,
\begin{equation}
\hat{S}=\ket{0}\bra{0}\otimes\sum_{x\in\mathbb{Z}}\ket{x-1}\bra{x}+\ket{1}\bra{1} \otimes \sum_{x\in\mathbb{Z}}\ket{x+1}\bra{x}. 
\label{shift}
\end{equation}
The combination  of both the  coin and shift  operators, 
denoted $W  \equiv \hat{S}(\hat{C}\otimes  \mathbb{I}_P)$, represents
the  walk   operator  on  the  total   Hilbert  space. Consequently, the  quantum state  of the particle  after $t$  steps of
noiseless quantum walk is the  linear superposition of $2t+1$ position
states,
\begin{equation}
\ket{\psi(t)} = \hat{W}^t\ket{\psi(0)} 
\label{1Rwalkop}
\end{equation}
where $W^t$ denotes  applying the walk operator ($W$)  $t$ times.  The
noisy  case  is  handled  numerically, in  the  conventional  way,  by
applying an instance of the noise super-operator $\mathcal{E}_t
\equiv \mathcal{E}(t+1,t)$ 
(realized by the Kraus  operators $K_j(t+1:t)$ after each application
of                               the                               $W$
operation \cite{peng2013non,chandrashekar2007symmetries,banerjee2008symmetry},
where  the Kraus  operators must  correspond to  the intermediate  map
(time   evolution   from   $t$   to  $t+1$).   In   this   case,   Eq.
(\ref{walkop}) is replaced by:
\begin{equation}
\rho(t) = \bigg[\Pi_t (\mathcal{E}_t\mathcal{W})\bigg](\rho(0))
\label{eq:walkop0}
\end{equation}
where $\rho(0) \equiv \ket{\psi(0)}\bra{\psi(0)}$ and 
$\mathcal{W}$ is the walk superoperator corresponding to $W$ (i.e, the
action $W \rho W^\dag$)  i.e., action of unitary $W$ on a density operator $\rho$, while $\Pi_t$  is the product of the time sequence of operations: $({\mathcal E_t}{\mathcal W}) ({\mathcal E_{t-1}}{\mathcal W})  ... ({\mathcal E_1}{\mathcal W})$.

With  non-Markovian   noise,  especial  care  is   needed,  since  the
intermediate   map  could   be   a   non-completely  positive   (NCP).
Importantly, the backflow  element of RTN, to be  studied below, means
that this  map may cause  the Kraus operator-sum representation  to be
replaced  by  an  operator-sum-difference representation,  a  detailed
discussion of this will appear in  section III.  In this present work,
we adopt a simpler approach,  where the Kraus operators $K_j(0;t)$ are
applied once after $t$  QW steps \cite{Banerjee2017NMQW}.  One expects
this to qualitatively reproduce the significant features that would be
obtained under the  application of the intermediate map  at each step.
We discuss elsewhere the difference between these two methods.

\subsection{Non-Markovian Noise}
Here we briefly discuss, from the perspective of their stochastic properties, local non-Markovian dephasing noise, due to the environment, which will be subsequently acted on the coin degrees of freedom of the quantum walker. 

\subsubsection{Random Telegraph Noise}
We now  discuss  a  local non-Markovian  dephasing  noise,  studied
in \cite{daffer2004depolarizing} and motivated by the Random Telegraph
Noise  (RTN) \cite{rice1944}.  The effect  of RTN  on the  dynamics of
quantum  systems, specifically  quantum correlations  such as  quantum
discord          has         been          studied         extensively
in  \cite{mazzola2011frozen,ali2014dynamics,pinto2013sudden}. RTN  was
also studied in the context of control of open system dynamics
\cite{control}.

The autocorrelation function for the RTN, represented by the stochastic variable $\Upsilon(t)$, is given by,
\begin{equation}
\langle \Upsilon(t) \Upsilon(s) \rangle = a^2 e^{-|t-s|\gamma}, 
\label{AC}
\end{equation} 
where $a$ has the significance of the strength of the system-environment coupling and $\gamma$ is proportional to the fluctuation rate of the RTN. The corresponding  power spectral density is a Lorentzian with peak value given by $2a^2/\gamma$. The Kraus operators, representing the RTN dynamical map are given by,

\begin{eqnarray} 
K_1(t) &= \sqrt{\frac{1+\Lambda(t)}{2}} I, \nonumber \\
K_2(t) &= \sqrt{\frac{1-\Lambda(t)}{2}} \sigma_3,  \label{krauss_rtn}
\end{eqnarray}

satisfying the completeness relation $\sum_{n=1}^{2}K_n^\dag{K_n}=I$.  $\Lambda(t)$ is the noise function based on the damped harmonic oscillator model that encomapsses both the Markovian and non-Markovian limits of the noise acting on the qubit,
\begin{eqnarray}
\Lambda(t) = e^{-\gamma t}\left[\textmd{cos}\left(\left[\sqrt{(\frac{2a}{\gamma})^2-1}\right]\gamma t \right)+
\frac{\textmd{sin}\left(\left[\sqrt{(\frac{2a}{\gamma})^2-1}\right]\gamma t\right)}{\sqrt{(\frac{2a}{\gamma})^2-1}}\right],
\label{kernal}
\end{eqnarray}

where  $\sqrt{(\frac{2a}{\gamma})^2-1}$   is  the  frequency   of  the
harmonic oscillators.  A power series expansion of $\Lambda(t)$ indicates  the absence of the term linear in $t$ bringing about a fundamental difference between white noise and colored noise.
The  function $\Lambda(t)$  corresponds to  two regimes;  the purely
damping  regime,  where  $a/\gamma<0.25$,  and  damped  oscillations  for
$a/\gamma>0.25$. Corresponding to these two regimes of $\Lambda(t)$, we
observe   Markovian   and    non-Markovian   behavior,   respectively,
  both  according  to the  CP-divisibility (Fig. (2)) and  backflow criteria (Sec. (4 and 5)). 

\begin{figure*}
	\centering 
	\includegraphics[scale=0.5]{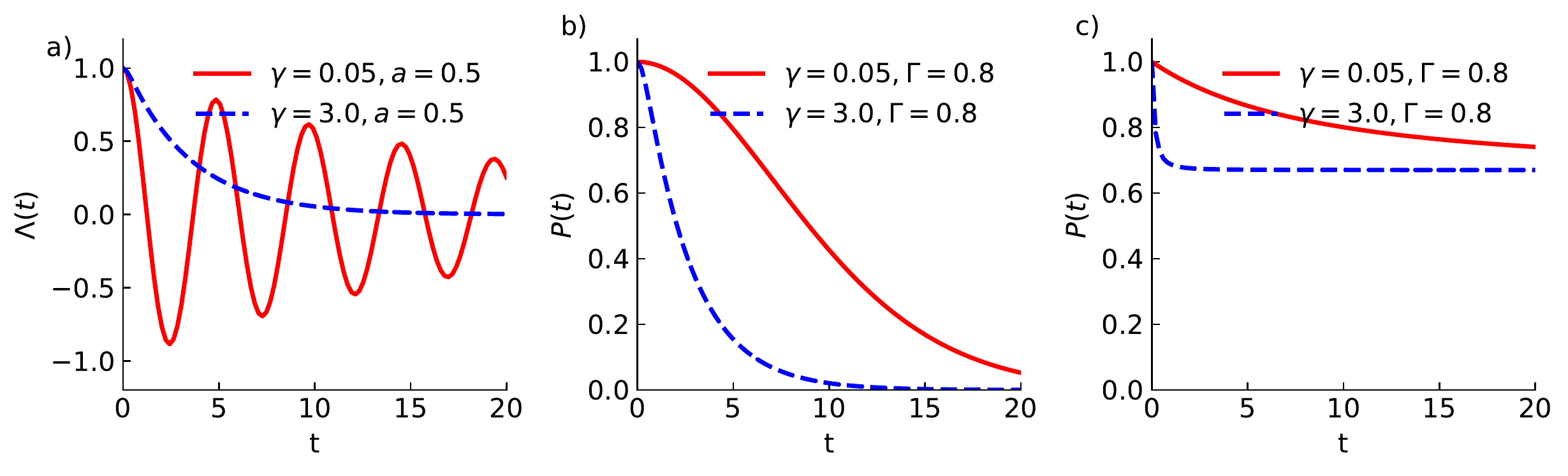}
	\caption{(Color  online) Plot of the noise functions that appear in the noise models. Note that the blue dashed line represents the Markovian regime while the red solid line corresponds to the non-Markovian regime. (a) RTN: The damped oscillations curve corresponds to the non-Markovian regime of the noise while the exponential decay curve is the typical Markovian regime (b) OUN: Unlike RTN, the non-Markoivan regime of OUN does not show damped oscillations. Instead, it has a power law behaviour which in contrast to the exponential decay observed in the Markovian regime. PLN: Similar to OUN, P(t) decays without any oscillations.}
	\label{MK}
\end{figure*}

\subsubsection{Modified Ornstein-Uhlenbeck and Power law Noises}

Here, we  briefly introduce the modified  Ornstein-Uhlenbeck and Power
law  noises, which,  as we  will see,  have similar  structures.  The
modified Ornstein-Uhlenbeck noise, from now  on referred to as OUN, is
a well known stationary Gaussian random  process which is in general a
non-Markovian process but possess a well defined Markov limit
\cite{uhlenbeck1930theory}.  An  example of  a physical  scenario that
leads to  the OUN  is when the  spin of an  electron interacts  with a magnetic  field subject  to  stochastic  fluctuations.\par

 Power law noise (PLN), also called $1/f^\alpha$ noise is
a non-Markovian stationary noise process,  where $\alpha$ is some real
number.   The  name  `Power  Law'  is  attributed  to  the  functional
relationship between  the spectral  density and  the frequency  of the
noise. The PLN is a major source of decoherence in solid state quantum
information    processing    devices     such    as    superconducting
qubits  \cite{paladino20141/f}.   PLN,  like  OUN,  is  in  general  a
non-Markovian  process but  possess a  well defined  Markovian limit.
\par

	The autocorrelation functions for OUN and PLN, represented by the stochastic variable $\Omega(t)$, are
	\begin{equation}
	\langle \Omega (t) \Omega(s) \rangle=
	\begin{cases}
	 \Gamma \gamma  e^{-\gamma|t-s|},~~~~~~~~~~~~~~~~~~~\textrm{OUN}\\\\
	  \frac{1}{2}(\alpha-1)\alpha\Gamma\frac{1}{(\gamma|t-s|+1)^\alpha}.~~~~\textrm{PLN}
	\end{cases}
	\end{equation}
The spectral properties of the noises are given by $\gamma$, which specifies the noise bandwidth and $\gamma^{-1}$ = $\tau_c$ is the finite correlation time of the environment. The parameter $\Gamma$ is the effective relaxation time which is an experimentally determinable quantity, often referred to as $T_1$ time in the literature. 

The dynamical map corresponding to the system density matrix can be expressed in terms of the Kraus operators as \cite{yu2010entanglement} 
 \begin{eqnarray}
K_1(t) &=  \sqrt{\frac{1+P(t)}{2}} I,\nonumber \\ 
K_2(t) &= \sqrt{\frac{1-P(t)}{2}}\sigma_3,
\label{krauss_ou}
\end{eqnarray}
where 

\begin{equation}
P(t)= \left\{
\begin{array}{ll}
\textrm{exp}[-\frac{\Gamma}{2} (t+\frac{1}{\gamma}(e^{-\gamma t}-1))],
~~~~~~~~~ & \textrm{OUN}\\ 
~ & ~ \\
\textrm{exp}(-\frac{t(t\gamma+2)\Gamma\gamma}{2(t\gamma+1)^2}) 
& \textrm{PLN}.
\end{array} \right.
\label{1Req:poft}
\end{equation}

 All the parameters have the same meaning as above. 

For the OUN scenario, for sufficiently large values  of the noise bandwidth, $\gamma
\rightarrow \infty$, the  correlation time $\tau_c \rightarrow
0$    and    hence    the   correlation    function    becomes
$\langle\Omega(t) \Omega(s) \rangle   \rightarrow   \Gamma   \delta(t-s)$,
whence the  Markov property  of the OUN  is obtained.  In this
limit,  the  coherence  terms  decay exponentially as  $P(t)  \rightarrow
e^{-\Gamma t}$.  On  the other hand, when  $\gamma$ is finite,
the resulting spectrum  is Lorentzian as opposed  to the delta
function in  the Markovian  regime.  Here,  the decay  rate slows
down to a polynomial value of $P(t) \rightarrow 1-\frac{1}{2} \Gamma \gamma t^2$ (in the limit $\gamma \approx 0$). In Fig. \ref{MK}, different regimes of all the noise sources considered are explicitly shown. 

\section{Intermediate maps and complete positivity:}

Two major (often inequivalent) criteria for witnessing non-Markovian behavior are information backflow \cite{breuer2009measure,breuer2016colloquium} and departure from  CP-divisibility of the intermediate  map connecting the density  operators  $\rho(t_2)$ and  $\rho(t_1)$  at  times $t_2$  and $t_1$, with  $t_2>t_1$ \cite{RHP10,RHP14}. Both methods  are discussed here and found to be equivalent for the noise models considered.

Given a dynamical map $\mathcal{E}(t_2,t_0)$ linking a system's
density operator at times $t_0$ and $t_2 > t_0$, 
the intermediate map $\mathcal{E}^{\rm IM}(t_2,t_1)$
for some intermediate time $t_1$ such that  
$t_2 > t_1 > t_0$, is given by: 
\begin{equation}
\mathcal{E}^{\rm IM}(t_2, t_1) = \mathcal{E}(t_2, t_0) \mathcal{E}^{-1}(t_1, t_0). 
\label{eq:inter}
\end{equation}
provided that the inverse map
$\mathcal{E}^{-1}(t_1, t_0)$ exists. 
The Choi matrix for the intermediate map can be obtained as:
\begin{eqnarray}
\textrm{M}_{\rm choi} = 
(\mathcal{E}^{\rm IM}(t_2,t_1) \otimes \mathbb{I})\ket{\Phi^+}\bra{\Phi^+},
\label{eq:mchoi}
\end{eqnarray}
where $\ket{\Phi^+} \equiv \ket{00}+\ket{11}$.
In the following subsections,  following  the method  presented in  \cite{RHP14}, we briefly derive the intermediate dynamical  maps of   the    noisy   channels  RTN, OUN and PLN.

\begin{figure*}
	\includegraphics[width=\linewidth]{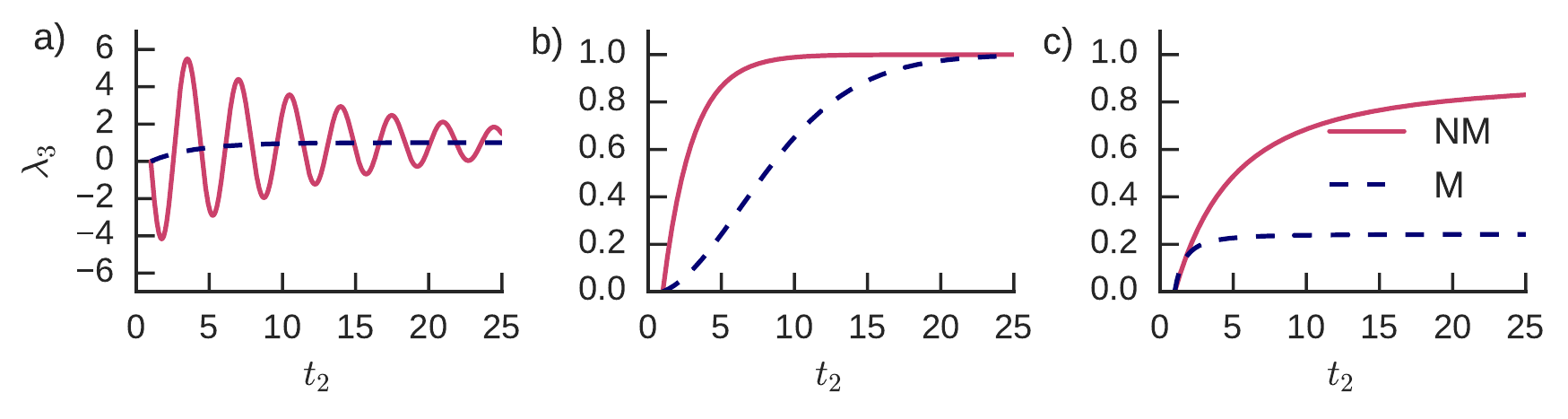}    
	 \caption{(Color  online)
	 The  eigenvalue $\lambda_3$  of the  Choi matrix
	obtained from intermediate dynamical map in both the Markovian
	and  non-Markovian  for  different   noise  models  (see  Eqs.
	(\ref{eq:l3rtn}) and (\ref{eq:l3pln})). The inital time $t_1 =
	1$ is  fixed for all the  plots (a) RTN: In  the non-Markovian
	regime  ($\gamma =  0.05,  a =  0.9$)  we observe  $\lambda_3$
	oscillating between  positive and negative eigenvalues  and in
	the Markovian  regime ($\gamma =  5, a = 0.9$)  $\lambda_3$ is
	always positive. (b) OUN: In both the non-Markovian ($\gamma =
	0.05 , \Gamma = 1.0$) and  the Markovian regime ($\gamma = 5.0
	, \Gamma =  1.0 $) $\lambda_3$ is positive.   (c) PLN: Similar
	to OUN, no  negative value for for $\lambda_3$  is ovserved in
	both  the non-Markovian  regime  ($\gamma =  0.05  , \Gamma  =
	5.0$),   Markovian  regime   ($\gamma   =  2.0   ,  \Gamma   =
	5.0$).  \label{fig:rhp}}
\end{figure*}

\subsection{Intermediate map for RTN}
The Choi matrix corresponding to the intermediate map 
for RTN is found to be,
\begin{eqnarray}
\textmd M_{\rm choi} = \begin{bmatrix}
1 & 0 & 0 & \frac{\Lambda(t_2)}{\Lambda(t_1)} \\
0 & 0 & 0 & 0 \\
0 & 0 & 0 & 0 \\
\frac{\Lambda(t_2)}{\Lambda(t_1)} & 0 & 0 & 1
\end{bmatrix}.
\label{eq:rtnmchoi}
\end{eqnarray}
The non-vanishing eigenvalues of $\textmd M_{\rm choi}$ in Eq. (\ref{eq:rtnmchoi})
are,
\begin{equation}
\lambda_3 = \bigg[1 - \frac{\Lambda(t_2)}{\Lambda(t_1)}\bigg] ; 
\lambda_4 = \bigg[1 + \frac{\Lambda(t_2)}{\Lambda(t_1)}\bigg].
\label{1Req:l3rtn}
\end{equation}
Here $\lambda_1$ and $\lambda_2$ are the vanishing eigenvalues. It
may  be  checked  that  if  $2a\ll \gamma$,  then  $\Lambda(t)$  is  a
monotonically decreasing function and  all eigenvalues are positive at
all times (Fig. \ref{fig:rhp}. a), consistent with the Markovian limit of RTN.  On  the  other hand,  if
$2a\gg \gamma$,  then $\Lambda(t)$ can  have regions of  increase, and
correspondingly,   some   eigenvalues    can   be   negative   (Fig.
\ref{fig:rhp}.a), consistent  with the non-Markovian regime  limit of the dynamical map.  

Generalizing    the     Choi    method    \cite{leung2003choi,choi1975completely} to  possibly NCP  maps, the Kraus  operators for
the   intermediate  map   are  obtained   by  ``folding''   the  above
eigenvectors \cite{OmkarSingleQubitChannels}
\begin{equation}
K_\pm^{\rm IM} = 
\sqrt{\frac{1}{2}\bigg|1 \pm \frac{\Lambda(t_2)}{\Lambda(t_1)}\bigg|}
\bigg( \begin{array}{cc}1&0\\0&\pm1\end{array}\bigg).
\label{eq:RTNIM}
\end{equation}
Note that if all eigenvalues  are positive, then these Kraus operators
satisfy  the  completeness  condition   $K_+^{\rm  IM \dagger}K_+^{\rm  IM}  +
K_-^{\rm  IM \dagger}K_-^{\rm IM}  =  \mathbb{I}$ and  the  evolution has  the
operator-sum  form: $\rho  \longrightarrow  \sum_j  K_j^{\rm IM}  \rho
K_j^{\rm IM\dag}.$  This characterizes the situation  in the Markovian
regime.

But  if some  eigenvalues are  negative (in  fact, $\lambda_3$),  then
observe that because of the way the absolute value is taken in defining
the  Kraus   operators,  these  operators  satisfy   the  completeness
condition with a minus sign. Quite generally, we have
completeness of the type
\begin{equation}
K_-^{\rm IM \dagger}(t_1,t_2)K_-^{\rm IM}(t_1,t_2) \pm K_+^{\rm IM \dagger}(t_1,t_2)
K_+^{\rm IM}(t_1,t_2)  =   \mathbb{I},
\end{equation}
where the  sign $\pm$ is determined  by whether or not  $\lambda_3$ is
negative.   Correspondingly, the  evolution  has  the operator-sum  or
operator-sum-difference form \cite{omkar2015operator}:
\begin{align}
\rho    &\longrightarrow   K_-^{\rm    IM}(t_1,t_2)   \rho    K_-^{\rm
	IM\dag}(t_1,t_2)    \pm    K_+^{\rm   IM}(t_1,t_2)    \rho    K_+^{\rm
	IM\dag}(t_1,t_2)\nonumber\\    &\equiv   \mathcal{E}(t_1,t_2)[\rho].
\label{eq:alyn}
\end{align}
This  characterizes  the  general  situation, no  matter  whether  the
evolution is Markovian or non-Markovian.

\subsection{Intermediate map for OUN  and PLN}
Following the recipe (\ref{eq:mchoi}), the Choi matrix for OUN is obtained as
\begin{equation}
\textmd M_{\rm choi} = \begin{bmatrix}
1 & 0 & 0 & \frac{p(t_2)}{p(t_1)} \\
0 & 0 & 0 & 0 \\
0 & 0 & 0 & 0 \\
\frac{p(t_2)}{p(t_1)} & 0 & 0 & 1
\end{bmatrix},
\end{equation}
for which the eigenvalues are,
\begin{equation}
\lambda_1 = \lambda_2 = 0; \lambda_3 = \bigg[1 - \frac{p(t_2)}{p(t_1)}\bigg] ; \lambda_4 = \bigg[1 + \frac{p(t_2)}{p(t_1)}\bigg],
\label{1Req:l3pln}
\end{equation}
with the intermediate map Kraus operators given by:
\begin{equation}
K_\pm^{\rm IM} = 
\sqrt{\frac{1}{2}\bigg|1 \pm \frac{p(t_2)}{p(t_1)}\bigg|}
\bigg( \begin{array}{cc}1&0\\0&\pm1\end{array}\bigg).
\label{eq:OUNIM}
\end{equation}
 The parameter  $p(t)$ for the OUN and PLN  noises can be
seen from  (\ref{eq:poft}).
It  is evident  from the  form of  the probability  $p(t)$,  that this  function is  monotonically decreasing
and hence these eigenvalues are never negative. Therefore, even in the
non-Markovian regime, the intermediate map remains CP. Later, we shall
find that, similarly, no backflow  features show up in this case.

\begin{figure*}
	\centering 
	\includegraphics[scale=0.5]{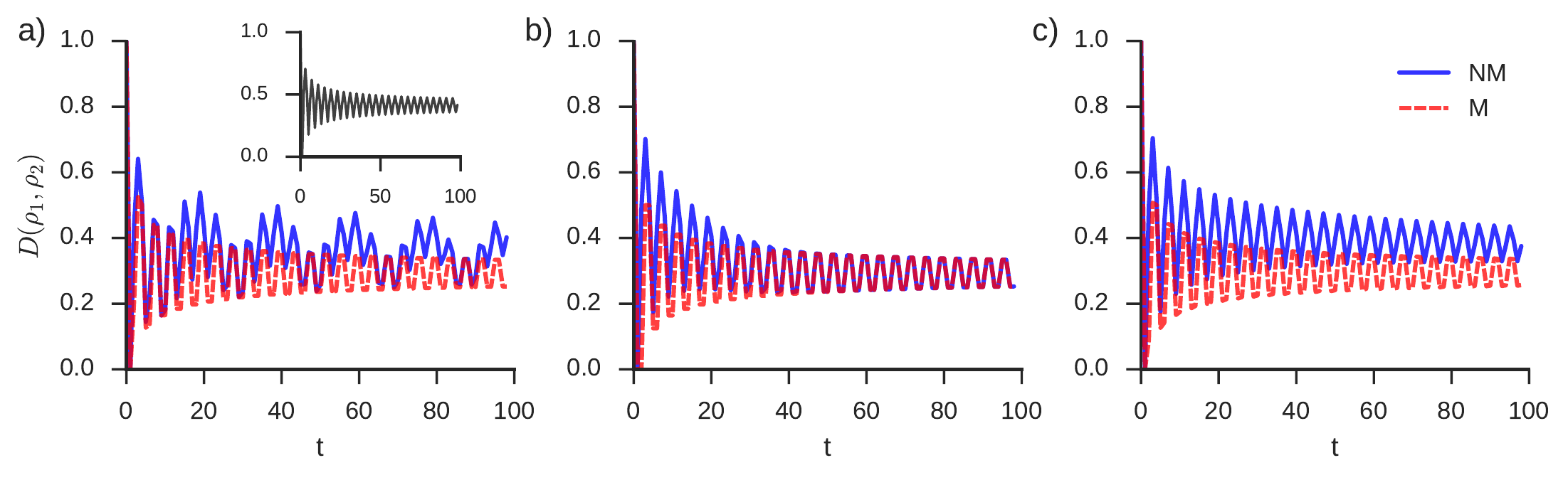}
	\caption{(Color  online)  Plot  of  TD  evolution with respect to number of time steps $t$, for the initial states $\ket{\psi(\pm\frac{\pi}{4}, 0)}$ under  the influence of different non-Markovian noise sources. The curves in the plot indicate the noiseless quantum walk (inset plot), the Markovian regime (red dashed line) and non-Markovian  regime (blue solid line) of the respective noise sources.  (a)  RTN:  The  noiseless quantum walk, i.e., the QW  in the  absence of  an external  noise, shows rapid  recurrences  due  to interaction  with  the  position ``environment", while the non-Markovian regime ($\gamma = 0.001, a = 0.08$) shows an  additional oscillatory term attributed to the non-Markovianity  in  the  environment-induced  decoherence, which is absent in the Markovian regime ($\gamma = 7.0, a = 1.0$). (b) OUN: In the Non-Markovian  regime ($\gamma = 0.01, \Gamma = 0.1$), TD decays without the additional recurrent feature  seen in the RTN  case. (c) PLN is similar to OUN in this respect. Noise parameters are same as OUN}
	\label{fig:TD}
\end{figure*}

Below,  we  shall  consider non-Markovianity  indicated  by  the
distinguishability criterion  applied to  the reduced dynamics  of the
coin in a DTQW, subjected to the above three noise models. 

\section{Signatures of non-Markovianity}
It is evident  from the above discussions  that non-Markovian behavior
is  capable  of throwing  a  number  of  surprises. Hence  it  becomes
imperative  to  have means  by  which  this  behavior can  be  studied
quantitatively. In what  follows, we will attempt this  using two well
known measures,  trace distance  (TD) between two  initial states,
and  coin-position  mutual  information (MI)  to  study  non-Markovian
backflow behavior under QW evolution.

\subsection{Trace distance}
As mentioned before, a powerful diagnostic of non-Markovian character, encoded in the dynamics of quantum systems,  is the backflow of information from the environment to the system. This, in turn manifests in the form of oscillations in correlation measures such as quantum mutual information \cite{luo2012quantifying}, and trace distance \cite{laine2010measure}, which are otherwise monotonic functions if the dynamics is Markovian. The distance between any two quantum states defined on the space of density matrices is given by a metric called the TD. It is defined as    
\begin{eqnarray}
D(\rho_1,\rho_2) = \frac{1}{2}\rm{Tr} \norm{\rho_1-\rho_2},
\label{td}
\end{eqnarray}
where $\norm{A}$ is the operator norm given by $\sqrt{A^{\dag} A}$. TD encompasses the idea of distinguishability of a pair of quantum states, which is monotonically decreasing under completely positive (CP) maps $\Xi$, $\textit{i.e.,}$ the CP maps are contractions for this metric,
\begin{equation}
D(\Xi \ \rho_1,\Xi \ \rho_2) \leq D(\rho_1,\rho_2).
\end{equation} 

For non-Markovian  processes, due  to the backflow  of information
from the environment to the  system, there is a temporary increase
in the  distinguishability of quantum  states and hence  the above
inequality may be violated (this being the characteristic of
backflow).   This idea  has  been exploited  in  an effort  to
quantify non-Markovianity \cite{breuer2009measure}.

Initially,  we  consider  the  noiseless (unitary)  evolution  of  the
quantum walk.  To study  the reduced  dynamics of  the coin  state, we
initialize the  quantum states, $\ket{\pm}  = \frac{\ket{0}\pm\ket{1}}
{\sqrt{2}}$, and compute TD between the two states defined above using
Eq. (\ref{td}).  Since the evolution  is governed by  unitary dynamics
the overall  evolution of the  quantum walk remains unitary  and hence
the TD  is preserved.   However, TD measure  between the  reduced coin
states   undergo  high   frequency  oscillation   as  shown   in  Fig.
\ref{fig:TD}.  Such  TD oscillations are a  signature of non-Markovian
backflow  behavior  leading to  non-zero  value  of the  non-Markovian
measures  \cite{Hinarejos2014}.   In   addition  to  the  oscillations
present  in  the  noiseless  evolution  of  quantum  walk,  a  further
oscillatory or recurrence structure arises when the coin is exposed to
an external noise such as RTN in the non-Markovian regime, as depicted
in Fig. \ref{fig:TD}.(a).  We  also observe that unlike RTN, neither
OUN  nor PLN  causes  new  recurrences. Instead,  non-Markovianity
manifests by  way of the  trace distance  between the two  coin states
diminishing more slowly as compared to their  Markovian counterpart, as shown in
Figs. \ref{fig:TD}.(b) and (c), respectively.

\begin{figure*}
	\centering
	\includegraphics[scale = 0.48]{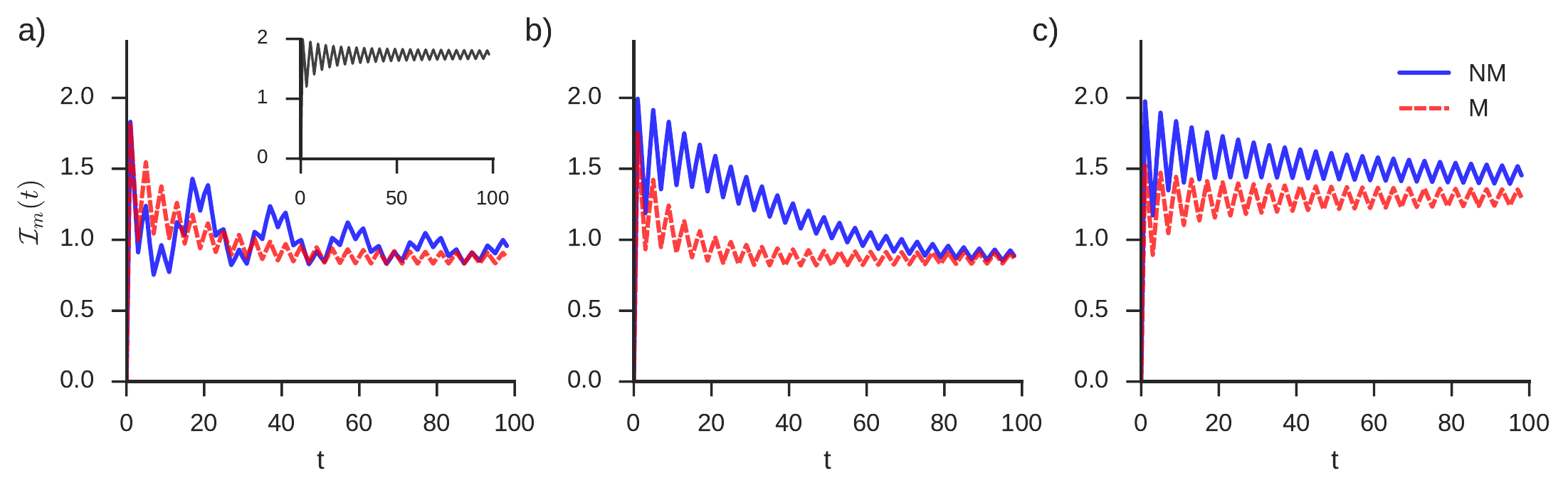}
	\caption{(Color online) Plot of MI evolution with respect
		to  number  of  time  steps  $t$, $\ket{\psi(\pm\frac{\pi}{4}, 0)}$, under  the  influence  of
		different  non-Markovian noise  sources. The  curves in  the
		plot indicate the noiseless quantum walk (inset),
		the  Markovian regime  (red dashed  line) and  non-Markovian
		regime   (blue  solid   line)   of   the  respective   noise
		sources. (a) RTN: Oscillations are observed in the noiseless
		quantum  walk,  similar to  TD,  arising  from the  position
		environment. In the non-Markovian regime ($\gamma = 0.01, a
		= 0.08$),  the information backflow  arising in the  form of
		oscillations  shortly  after a  few  initial  time steps  is
		observed. (b)  OUN: In  the Non-Markovian regime ($\gamma = 0.01, \Gamma=0.5$),  MI decays
		without any recurrences from the environment, indicating the
		absence of the  backflow feature which manifests  in the RTN
		case.   However  the   correlation  is   preserved  in   the
		non-Markovian  regime  for longer  time  steps.  (c) PLN  is
		similar to OUN but preserves the correlation for a longer time
		in both the regimes. Noise parameters are same as OUN}
	\label{fig:MI}
\end{figure*}

\subsection{Mutual Information}
Quantum correlations has been used to  understand non-Markovianity in dynamical maps. In particular, MI has been used to quantify non-Markovianity \cite{luo2012quantifying}. Let $\rho_1$ and 	$\rho_2$ be the density matrices representing the system and the ancillary states, respectively. The 		quantum correlation between the $\rho_1$ and $\rho_2$ is given by, 
\begin{equation}
\mathcal{I}(\rho) = S(\rho_1)+S(\rho_2)-S(\rho), 
\end{equation}
where $S(.)$ is the Von Neumann entropy $S(\rho_1):= - \rm{Tr} \rho_1 log_{2}\rho_1$. Similar to the TD 	measure, MI is also a monotonically decreasing function when the dynamics is Markovian, 
\begin{equation}
\mathcal{I}((\Xi \otimes \textmd{I})\rho) \leq \mathcal{I}(\rho), 
\label{MIm}
\end{equation}
leading to a measure of non-Markovianity based on MI. The  quantum correlation  between the  coin  and the  position can  be
quantified using  MI. This in turn  could serve as a  method to detect
non-Markovianity  in  the  quantum  walk  evolution.   Similar  to  TD
oscillation seen  in the noiseless  quantum walk, MI also  encodes the
same features, as shown in Fig. \ref{fig:MI}. Similar to the variance
of the  QW, MI is  useful for  demonstrating the quantum  to classical
transition. When  the coin is driven  by a noise, irrespective  of its
characteristics, MI tend to vanish in the long time limit. However, it
is  the short  time  behavior that  is  of interest  to  us here.   In
Figs. \ref{fig:MI}.(a-c), are depicted the  results of the  short time
behavior of MI  under different noise models. In the  case of RTN, due
to the presence  of backflow of information,  MI revives periodically,
as shown  in Fig. \ref{fig:MI}(a).  In  the case of OUN  and PLN,
non-Markovianity manifests  by way of preservation  of the correlation
for a longer period of time compared to the Markovian regime, as shown
in Fig. \ref{fig:MI}(b).  Now we discuss a method to disambiguate
different  sources of  non-Markovianity.  This  is natural  to the  QW
system, studied here,  but could be also be envisaged  in more complex
systems, such as  the suppression of decoherence due  to two competing
sources leading  to frustration,  a feature of  use in  fault tolerant
holonomic quantum computation \cite{Affleck,GPFrustration}.

\section{Disambiguating different sources of non-Markovianity}

\begin{figure}
	\centering
	\includegraphics[scale = 0.7]{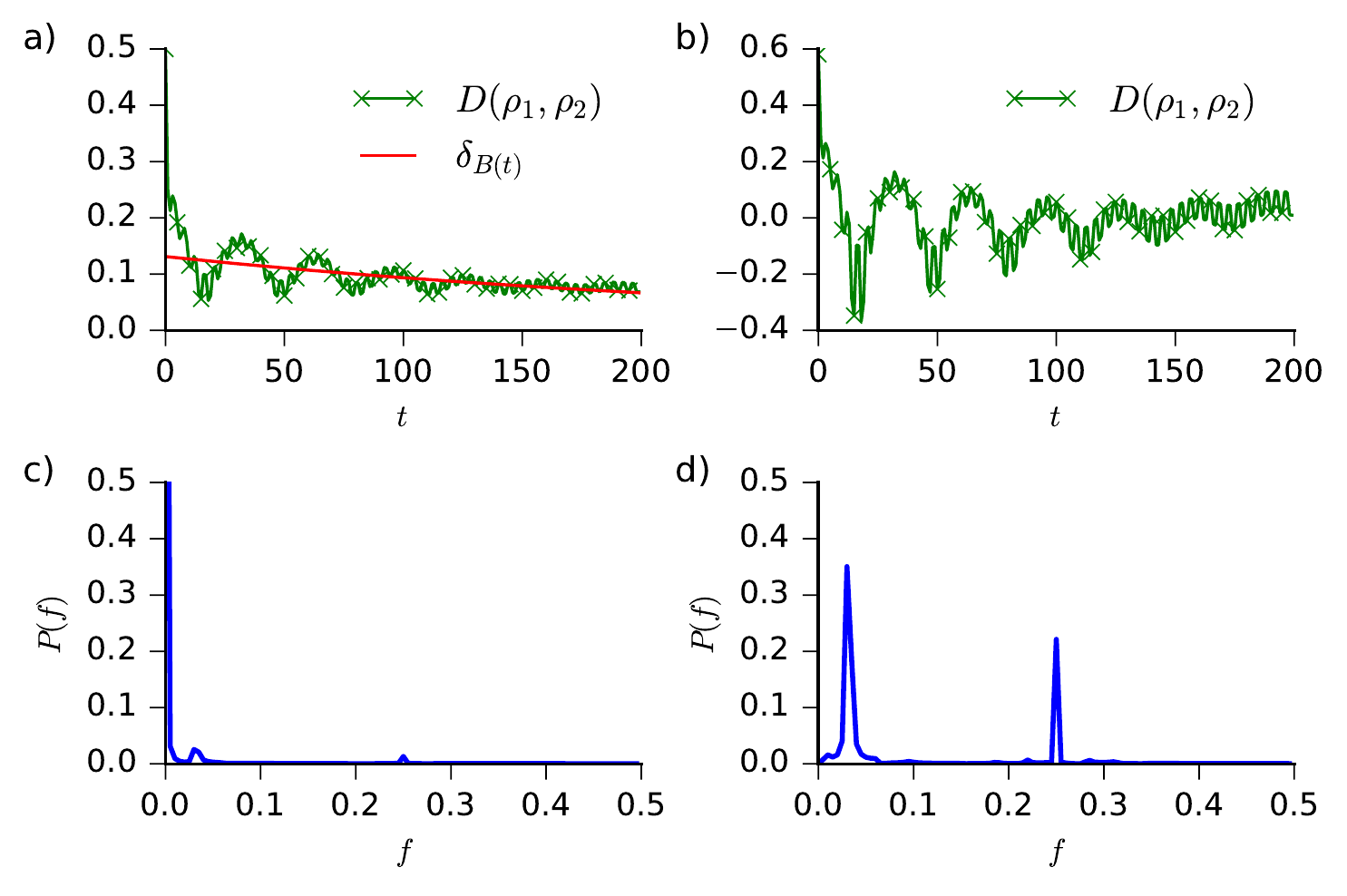}
	\caption{(Color  online)   Filtering  and  power
		spectral  analysis  of  TD  between coin  states $\ket{\psi(\pm\frac{\pi}{4}, \frac{\pi}{3})}$ driven  by
		RTN. (a) Plot of the actual TD curve (green solid
		line) and  the corresponding  best monotonic fit (black
		dashed line). (b) Filtered TD obtained by
		subtracting the  exponential best  fit from the  actual TD
		curve. (c) The actual  spectrum of the trace distance
		without  filtering with a large low frequency component. (d)  The power spectrum of the filtered TD,  shows two frequency peaks, the
		one around $f=0.02$ (resp.,  $f=0.25$) corresponding to
		the   RTN  (resp.,   position)  recurrence   component.  The
		importance of this  filtering approach is that  it allows us
		to  compare the  relative strengths  of the  two sources  of
		non-Markovianity.}
	\label{fig:spectrum}      
\end{figure}

The study of both TD and MI in the previous section, clearly shows how
non-Markovianity originates  from both the subsystem  dynamics such as
position and  purely from  environmental sources like  RTN. In  such a
scenario, identifying the different  sources of non-Markovianity would
be useful  in comparing  the relative  strengths of  different sources
with  respect to  one another.  Here, we  employ a  recently developed
\cite{Banerjee2017NMQW}  time  domain  filtering  and  power  spectral
method to disambiguate the  position non-Markovianity from environment
induced non-Markovianity.  Our method  is grounded on  the observation
that,   different  sources   of   non-Markovianity  induce   different
frequencies of oscillation in the time evolution of either TD or MI or
any other similar correlation measure.  By applying a proper filtering
function to TD or MI we  can detect the frequencies of the oscillation
induced       by      both       position      and       environmental
noise. Fig. \ref{fig:spectrum}.(c) is the  actual power spectrum of TD
between two coin states driven  by RTN and Fig. \ref{fig:spectrum}.(d)
is the  filtered power spectrum  obtained by subtracting the  TD curve
from its corresponding Monotonically Falling Best Fit Function (MFBF).

The MFBF  is simply a best fit function  subject to the condition of
strictly monotonic decrease. The form of the function is guided by the
form of the function in  the corresponding Markovian limit.  The basic
rationale  behind  this  approach   is  that  with  high  probability,
different  sources  of backflow  (e.g.,  the  environment and  another
subsystem,  such as  the  coin or  position)  correspond typically  to
different recurrence  time-scales overlaid on an  underlying Markovian
trend. This  trend can be filtered  out as a low  frequency component,
whereas the  various backflow  contributions can  be separated  in the
Fourier  domain  as different  frequency  components.  We stress  that
subtracting the Markovian  trend as the MFBF is  a heuristic principle
to  pre-process the  data before  Fourier analysis.   Future work  can
evolve more finely tuned ways to implement this idea.

The most  general recipe  for identifying  the MFBF can  be cast  as a
monotone regression  problem, in  which the best  fit function  can be
obtained  using  constrained  optimization  techniques.  For  example,
minimizing  non-linear  least  squares subjected  to  strict  monotone
decreasing  condition  amounts  to solving  a  quadratic  optimization
problem \cite{best1990MBF},
\begin{eqnarray}
&& {\rm minimize}\quad \sum_{i=1}^{n} w_i(g(x_i)-f(x_i))^2 \\
&& \mbox{\rm subject to}\quad  x_{i+1} - x_i < 0\,,  \label{constraint}
\end{eqnarray}
where $g(x_{i})$, $i=1,\ldots,n$, is  the best fit function  subjected to condition
(\ref{constraint}),   $f(x_{i})$, $i=1,\ldots,n$,  is   the   given  function   and
$w_{i}$, $i=1,\ldots,n$, is the appropriate weight.

The filtered  spectrum, as shown is obtained by fitting an exponential curve which serves as the MFBF, which when subtracted from the actual TD curve clearly  detects the  presence of both  position  and  environment induced  non-Markovianity.  In Fig. \ref{fig:spectrum}.(c) the  peak around f = 0.27 (resp. f  = 0.025) corresponds to the position (resp., RTN) non-Markovian source,  with the power associated  with the former about 1.5  times more.  The  key point here is  that in the  RTN case, information  backflow  to   the  system  from  the   position  or  RTN
reservoirs, show up as separate peak frequencies in the power spectrum
of Fig.  \ref{fig:spectrum}.(d).  However, for  the OU noise,  we find
that  information backflow,  indicated by  the secondary  peak in  the
filtered  power spectrum,  is absent.  Here, the  reservoir has  a low
bandwidth, $\gamma$,  resulting in a large  reservoir correlation time
in relation to the system correlation time.

\section{Quantumness of quantum walk}

The time evolution of  quantum walk induces non-classical correlations
between coin and  position degrees of freedom. The  usual signature of
quantum walk is  the quadratic growth in variance, in  contrast to the
classical    random    walk    for    which   it    is    linear    in
time \cite{ambainis2001one}  . Other  than studying  variance, greater
insight  into  the   non-Markovian  nature  of  the  QW   due  to  the
coin-position nonclassical  correlations can be developed  by studying
correlation  measures, such  as measurement-induced  disturbance (MID)
and Quantum Discord (QD)
\cite{SCSmid,SCSdiscord}.  In  addition,  we find  that the  quantum state purity  can be a  useful diagnostic here,  in the sense  that it indicates the effect of entanglement with environment. 

\begin{figure*}
	\includegraphics[width=\linewidth]{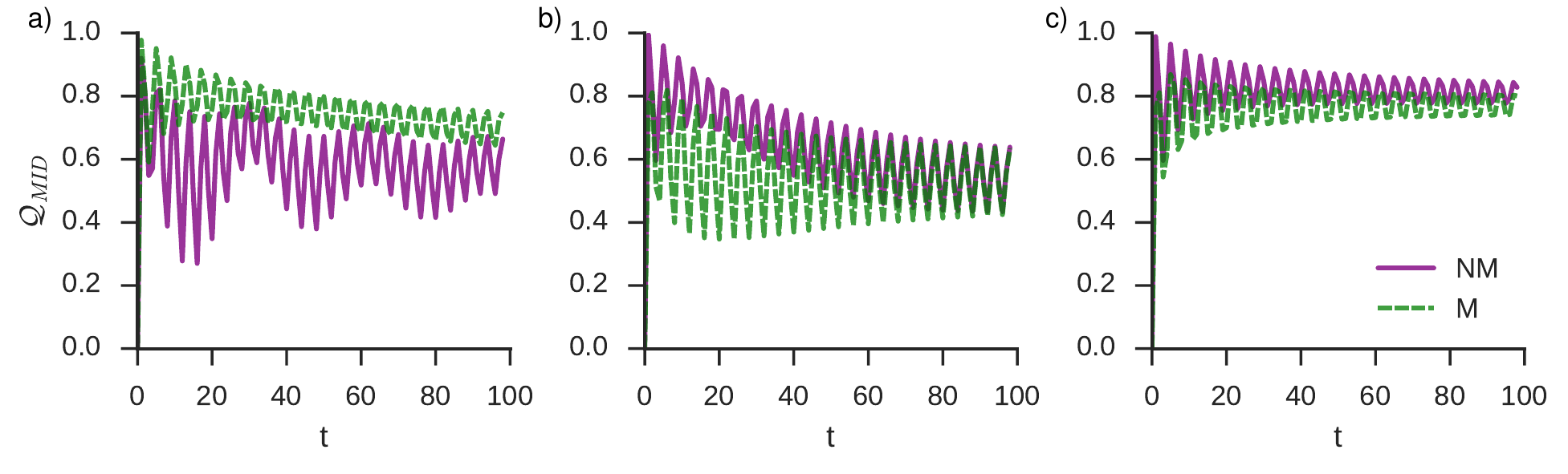}
	\caption{(Color online)  Plot of MID for  100 steps of
		quantum  walk for the initial state $\ket{\psi(\frac{\pi}{4}, 0)}$ in  the  presence  of  RTN,  OUN  and
		PLN. The non-Markovian regime is plotted as purple solid line and the Markovian regime is the green dashed line of the respective noise sources. (a) RTN: The interaction between
		coin   and   RTN   induces   oscillations   in   the
		non-Markovian regime  ($\gamma$ =  0.008, a  = 0.05)
		indicating  the  backflow of
		information from  the environment. In the Markovian regime ($\gamma$ =  2.0, a  = 0.05) backflow from RTN is absent and the oscillation is due to position induced non-Markovianity. (b) OUN:  In the
		Non-Markovian region  ($\gamma =  0.01, \Gamma=0.1$)
		unlike RTN no backflow due to the noise is observed, however the  decay  rate  is lesser compared to the Markovian
		regime ($\gamma$  = 2.0, a  = 0.1). (C). PLN is similar to OUN with no backflow in the non-Markovian regime. The trend observed 
		in the TD and MI cases, is also borne out here. }
	\label{fig:MID}
\end{figure*}

\subsection{Measurement Induced Disturbance}

To evaluate MID over the  QW evolution, we consider a state $\rho$ evolving in the Hilbert space $\mathcal{H}_c\otimes\mathcal{H}_p$ and the corresponding reduced density matrices represented by $\rho_c$ and $\rho_p$. Let the projectors $\left \{ \Pi_n \right \}$ denote the measurement operators, satisfying the usual properties such as completeness $\left \{ \sum_{n} \Pi_n= \textmd{I} \right \}$ and orthogonality $\left \{ \Pi_n \Pi_{n'}=\delta_{nn'}\Pi_n \right \}$. Here, $\Pi_n = \sum \Pi_c\otimes\Pi_p$,  where $\Pi_c$ and $\Pi_p$ denote the projective measurements on the coin and position density matrices $\rho_c$ and $\rho_p$, respectively. Thus, $\Pi_n$ is the joint projector (tensor product) acting on position space and coin space. It ranges from 1 to Dim(position)*Dim(coin). $\Pi $ is the super-operator acting on the bipartite density operator as a result on the above joint projective measurements.  The measurement induced state is given by \cite{luomid}
\begin{equation}
\Pi(\rho) = \sum_{i,j}(\Pi_c^i\otimes\Pi_p^j) \thinspace \rho \thinspace (\Pi_c^i\otimes\Pi_p^j).
\end{equation}

If the projectors that induces the state $\Pi(\rho)$ arise from the spectral resolution of the reduced density matrices, $i.e, \rho_c = \sum_{i}p_c^i \Pi_c^i$ and $\rho_p = \sum_{j}p_p^j \Pi_p^{j}$, then the marginal information content remains unchanged; in this sense, the measurement operators are attributed to be least disturbing or optimal. The measure of quantum correlations characterized by MID is the difference between the given density state $\rho$ and the measurement induced state $\Pi(\rho)$, 
\begin{equation}
\mathcal{Q}_{MID} = \mathcal{I}(\rho)-\mathcal{I}(\Pi(\rho)),
\end{equation}

where $\mathcal{I}(.)$ is the mutual information, defined as

\begin{eqnarray}
\mathcal{I}(\rho)=S(\rho_C)+S(\rho_P)-S(\rho) \label{Im}.
\end{eqnarray}

\begin{figure*}
	\includegraphics[width=\linewidth]{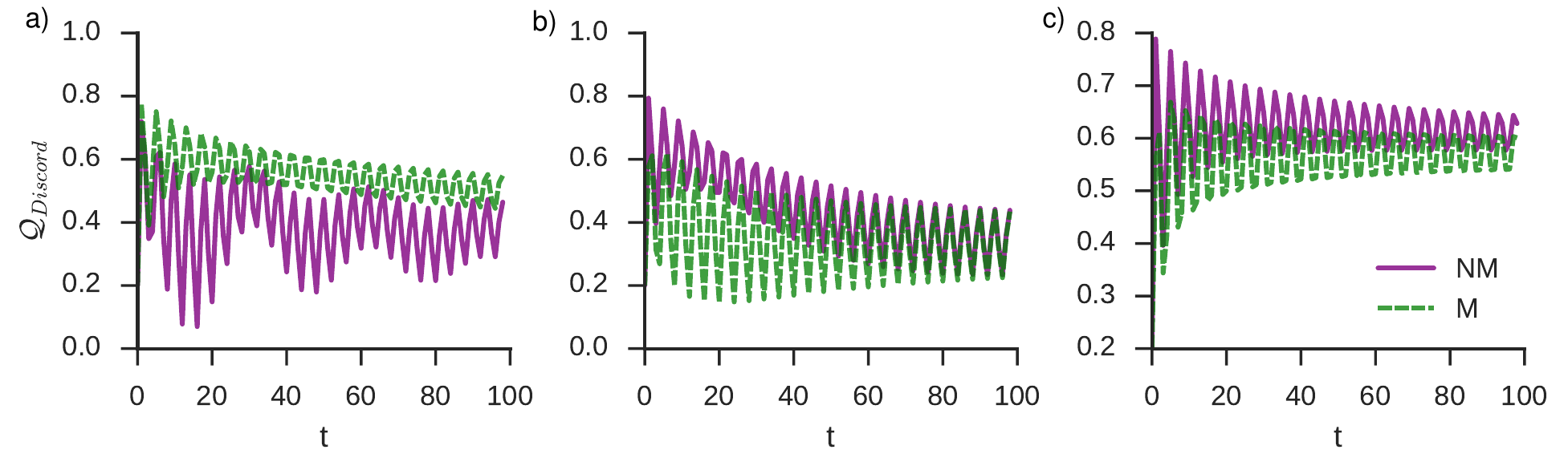}
	\caption{(Color online) Plot of  QD as a function of walk
		steps $t$,  under the  influence of RTN,  OUN and  PLN. Note
		that the noise parameters and the initial state of the particle are same as mentioned for MID. The
		results are similar  to MID for all three  noise sources (a)
		RTN,  (b)  OUN and  (c)  PLN,  respectively, that  drive  the
		coin. QD  is strictly a lower  bound on the MID  between the
		coin  and  the  position  states, due  to  the  optimization
		performed over the measurement operators. Once again, the trend  
		observed  in the TD and MI cases, is  borne out here.}
	\label{fig:discord}
\end{figure*}

\subsection{Quantum Discord}

Next,            we           use            quantum           discord
(QD)  \cite{ollivier2001discord,vedral2012discord}   to  estimate  the
difference  between the  total and  classical correlations  during the
course of  quantum walk  evolution.  QD is  defined as  the difference
between the two natural extensions  of classical mutual information to
the quantum  setting. The  first definition  of mutual  information in
given  by Eq.  (\ref{Im}) and  the  second definition  is the  quantum
version  of  conditional  entropy  which depends  on  the  measurement
process. It is defined as follows,
\begin{equation}
\mathcal{J}(\rho_p|\rho_c) = S(\rho_p) - S(\rho_p|\Pi_i^c),
\end{equation}
where $S(\rho_p|\Pi_i^c)$ is the quantum conditional entropy, defined with respect to a set of projective measurements $\left \{\Pi_i \right\}$ performed on the coin state $\rho_c$. The estimate of QD is then given by, 
\begin{equation}
\mathcal{D}(\rho_p|\rho_c)= \mathcal{I}(\rho) - \max_{\Pi^c_i} \mathcal{J}(\rho_p|\rho_c). \label{QD}
\end{equation}

QD  can be  thought of  as  an optimal  version  of MID.  In MID,  the
measurement  operators  that  characterize  the  quantum  correlations
between the coin and position  are given by the corresponding spectral
projections and  does not involve  any optimization. Hence, MID  is an
operational  measure in  the  sense that  computing  it is  relatively
simple. QD on  the other hand, optimizes over  all possible projectors
to compute the correlation between the coin and position states, which
makes it computationally intensive.

We have  computed both MID and  QD numerically for both  noiseless and
noisy   QW,   which   is    shown   in   Figs.   (\ref{fig:MID})   and
(\ref{fig:discord}), respectively. In both MID and QD oscillations are
observed in the non-Markovian regime of RTN. On the other hand OUN and
PLN  do  not  exhibit  any  oscillations  indicating  the  absence  of
backflow.    Note  that   the  oscillations  due  to  the
coin-position interaction vanishes rapidly, because the interaction of
the  coin with  the environment  makes the  full coin-position  system
increasingly mixed.

Given  that  quantum discord  is  not  symmetric, we  also  considered
$\mathcal{D}(\rho_c|\rho_p)$,   but  the   pattern   of  behavior   of
oscillations in the case of RTN, OUN and PLN, was similar to that seen
with  $\mathcal{D}(\rho_p|\rho_c)$, and  this figure  is not  included
here. 

\subsection{Quantum purity}

Quantum state purity is another interesting quantity that can be used to study the quantumness of quantum walks. Purity, similar to the correlation measures studied previously is a monotone under Markovian evolutions. However under non-Markovian evolutions temporary revivals in purity can be observed in the form of oscillations, this idea has been exploited and studied as a measure of non-Markovianity \cite{pati2017NM}. 

\begin{figure*}
	\includegraphics[width=\linewidth]{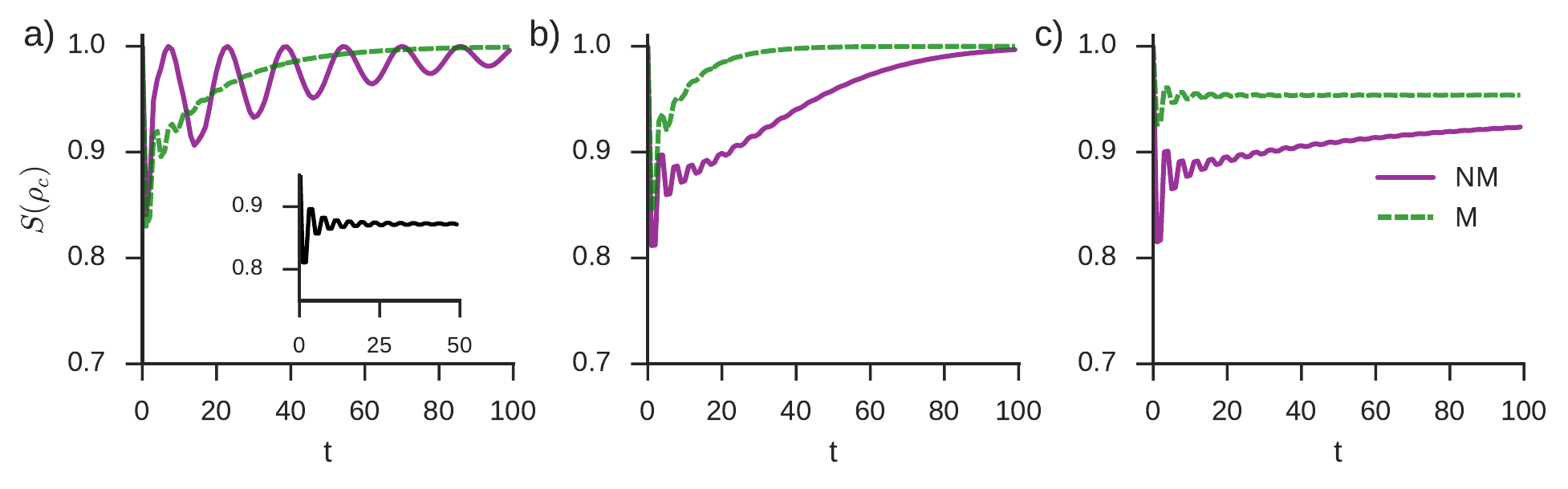}
	\caption{(Color online)Von-Neumann entropy of the state of the particle under different sources of non-Markovianity. (a) RTN: The backflow effect in the form of oscillations is clearly seen in the non-Markovian regime ($\gamma = 0.01, a = 0.1$). The inset plot is the von-Neumann entropy of noiseless quantum walk. (b) OUN: In both non-Markovian ($\gamma=0.01, \Gamma=1.0$) and Markovian regime ($\gamma=0.01, \Gamma=1.0$) only the position induced non-Markovianity can be observed. (C) PLN: Similar to OUN, the backflow effects are absent in PLN, the noise parameters are same as OUN. }
	\label{fig:ent}
\end{figure*}

Purity can be studied using Von-Neumann entropy $S(\rho_c) = - \rm{Tr} (\rho_c \ \textmd{ln} \ \rho_c)$ where $\rho_c$ is the reduced state of the particle obtained by tracing out the position degrees of freedom. Note that von-Neumann entropy is maximum for a completely mixed state for which the purity is zero. As noted in the previous discussions, due to the non-Markovianity induced because of the strong entanglement between the particle's internal and external degrees of freedom, the initially pure state of the particle is mapped to a mixed state ($S \approx 0.875$), this is shown in the inset plot in Fig. \ref{fig:ent} .(a). When the particle is coupled to external source of noise such as RTN, the state of the particle is driven to a maximally mixed state ($S = 1$). We notice that in the non-Markovian regime there is a temporary increase in the purity (decrease in entropy) of the state, this is shown in Fig. \ref{fig:ent}.(a). In accordance with all other previously observed witnesses of non-Markovianity such as TD, MI, both OUN and PLN does not show any increase in purity and the only source of revival in purity is induced by the position environment. 

\section{Conclusion}

Quantum non-Markovianity has  become an important area  of research in
the  last few  years.  Here  we  undertake the  task of  understanding
non-Markovian dynamics,  in the context  of quantum walks,  a familiar
``work horse''  in quantum  information processing.   Random telegraph
noise (RTN) exhibits backflow, manifesting in terms of oscillations or
recurrences in both  trace-distance and mutual information  as well as
not-Complete-positivity of  the intermediate map.  Non-Markovian noise
channels that  lack evident backflow are  found to be the  modified OU
noise  and the  power-law  noise (PLN).  We  develop a  power-spectrum
technique,   that   can   be   adapted  to   the   trace-distance   or
mutual-information  or  any  other  witness methods  of  backflow,  to
disambiguate external vs internal sources of backflow.  In a QW, these
are, typically, the decoherence  inducing environment and the position
degree  of  freedom.    We  also  study  various   facets  of  quantum
correlations in the transition from quantum to classical random walks,
under  the  considered  non-Markovian   noise  models.  All  of  these
contribute   to  our   efforts   to  understand   various  facets   of
non-Markovian  behavior.

\section{Acknowledgments}  SB acknowledges  support by  the project
number  03(1369)/16/EMR-II   funded  by  Council  of   Scientific  and
Industrial Research, New Delhi.  RS  thanks DST-SERB, Govt.  of India,
for    financial     support    provided    through     the    project
EMR/2016/004019. The work of V.~J.   and F.~P.  is based upon research
supported  by  the South  African  Research  Chair Initiative  of  the
Department of Science and Technology and National Research Foundation.

\end{document}